# Record fast-cycling accelerator magnet based on high temperature superconductors

Henryk Piekarz[1], Steven Hays, Jamie Blowers, Bradley Claypool and Vladimir Shiltsev

*Fermi National Accelerator Laboratory, Batavia, Illinois 60510, USA*

Four decades ago development of high-current superconducting NbTi wire cables revolutionized the magnet technology for energy frontier accelerators, such as Tevatron, RHIC and LHC. The NbTi based magnets offered advantage of much higher fields $B$ and much lower electric wall plug power consumption if operated at 4.5 K but relatively small ramping rates $dB/dt \ll 0.1$ T/s. The need for the accelerators of high average beam power and high repetition rates have initiated studies of fast ramping SC magnets, but it was found the AC losses in the low-temperature superconductors preclude obtaining the rates in the excess of (1- 4) T/s. Here we report the first application of high-temperature superconductor magnet technology with substantially lower AC losses and report record high ramping rates of 12 T/s achieved in a prototype dual-aperture accelerator magnet.

Particle accelerators critically depend on development of high-field superconducting (SC) magnets [1,2,3] which allow to extend the energy reach and achieve desired cost and electric power efficiency of major physics facilities such as Tevatron at Fermilab [4] – the pioneering machne in operation from 1987 to 2011, Relativistic Heavy Ion Collider [5] (RHIC, 1999 – now) at Brookhaven National Laboratory, and Large Hadron Collider[6] (LHC) at CERN, Switzerland which operates since 2008. Note, that all these accelerators mostly operate in the regime of very slow beam energy ramp and their magnetic field ramping rates $dB/dt$ are very low, 0.03 – 0.07 T/s. Next generation facilities such as muon colliders [7,8], future circular colliders[9] and high-intensity proton synchrotrons for neutrino research [10,11,12,13] accelerators demand substantially faster cycles of beam acceleration that in turn require fast cycling accelerator magnets with $dB/dt$ of the order of tens to hundreds of T/s. Normal conducting magnets can provide such rates – for example the JPARC 3 GeV proton rapid cycling synchrotron (RCS) magnets operate with $dB/dt$ rates of 70 T/s [14] – but resistive power loss in the conductor and magnetization loss in the magnetic cores steel make them prohibitively power inefficient. Operation of the world's largest accelerator complex at CERN requires about 180 MW electric power and three smaller, low-energy normal conducting RCS's altogether boosting the proton energy beam from 50 MeV to 450 GeV consume more electric power than much larger 6500 GeV SC LHC collider ring [15]. Fast cycling SC magnets face great challenges due to the AC losses - energy dissipation in the conductor caused mostly by the magnetization of the superconducting filaments and due to coupling currents between the filaments in the strands. State-of-the-art cryogenic systems require 930 W of wall plug power to provide 1 W of cooling capacity for NbTi SC magnets at 1.8 K, and 230 W/W at 4.5 K [17] and that poses very stringent limits on the allowable AC losses in the low-temperature superconducting (LTS) magnet accelerators. To-date, the highest ramping rates achieved in the operational LTS accelerator magnets are about 4 T/s [18,19]. When comes to the accelerator magnet technology, the high temperature superconductors (HTS) [20] have triple advantage against the LTS based on the NbTi or $Nb_3Sn$ superconductor– (i) much higher critical current densities and fields, (ii) lower AC losses and (iii) higher operational temperatures. In this Letter we report the 12 T/s ramp rates achieved in a dual-bore accelerator HTS magnet prototype.

---

[1] Corresponding author, electronic mail: hpiekarz@fnal.gov

The AC losses in a SC magnet are proportional to the mass of the conductor and depend on total current $I$, frequency $f$, maximum field $B$, and temperature $T$. The physical mechanisms and scaling of the AC losses in HTS tapes are different for the magnetization losses and for the transport-current losses are discussed in detail in [21, 22]. Of importance for the magnet design is that high current density of the HTS superconductors allows to strongly minimize the mass of the conductor and that the AC losses are significantly enhanced by the magnetic field components perpendicular to the tape surface. For typical rapid cycling operation expected in future accelerators the inductive loss component due to self-fields induced by the AC transport current will dominate the AC power loss. The inductance $L$ scales as $N^2$ with the number of turns and the minimal one can be achieved with a single-turn cable but for the required field $B$ in the gap of the magnet very high current $I$ conductor may be needed as $I \sim B/L$. For example, 2 T field has been achieved in the 20 mm double-gap superferric DC magnet with 100 kA current in a NbTi-based single-turn power cable [23]. For the fast-cycling operation, the power supply voltage $V$ grows with $dI/dt$ and can be prohibitively large, therefore, to optimize technical feasibility and cost of such power supply some compromise between the magnet current and the magnet inductance is required.

The rapid-cycling magnet design developed, tested and reported below – see figure 1 - has three novel features: a) it uses the high-current density HTS conductor; b) the conductor is placed inside the steel core of the magnet such that the magnetic field in the conductor is minimal; and c) its two vertically aligned beam gaps are energized by one conductor. The choice of a 3-turn conductor allows to operate the magnet with three times lower current than needed for a single turn option for the same field in the gap:

$$B = \mu_0 \, I \, N / g \qquad (1),$$

(here $g$ the gap size, $\mu_0$ is magnetic permeability of the vacuum) at the expense of acceptable 9-fold increase of the inductance. In such arrangement – conceptually proposed in [9] – the magnetic fields in the upper and lower gaps are of the same value but of opposite polarities that makes it uniquely beneficial for simultaneous acceleration of two beams at once – either beams of opposite charge particles (e.g., electron and positrons, positive and negative muons, protons and antiprotons) circulating in the same direction in each of the gaps, or two beams of the same particles circulating in opposite directions. Also of importance for the particle acceleration application is that in such design, the ever existing particle beam losses and decay products which have lower energy than the primary beams will be bent out and away from the HTS conductor, thus, minimizing its highly undesirable heating and therefore, greatly easing the requirements for the particle collimation and radiation protection systems [24].

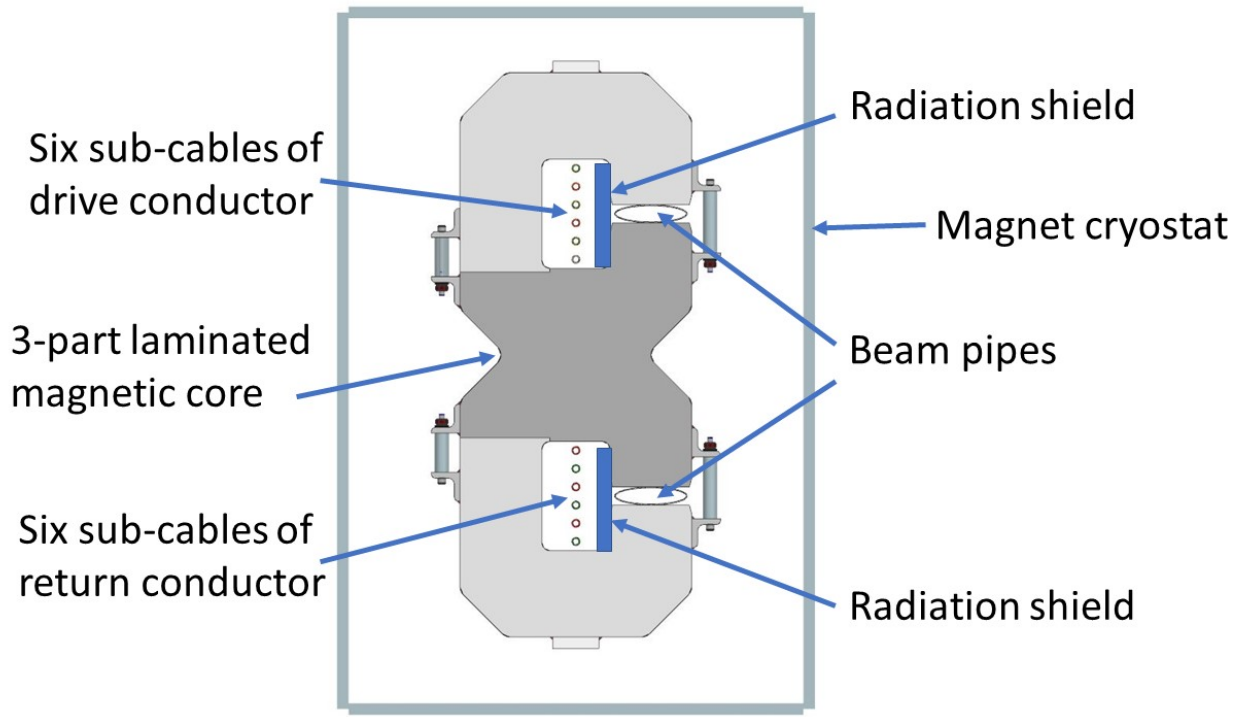

FIG. 1. A conceptual design of a vertical dual-bore HTS based accelerator magnet.

The quench propagation velocity in the HTS superconductor is very slow and that makes the quench detection and protection difficult [20, 25]. Operation of the HTS conductor at the temperatures much lower than the critical one allows efficient use of the temperature-based quench detection system. This is achieved by having the total cross-section of the HTS superconductor sufficiently large to carry the design transport current up to, e.g., 30 K, so at the operational temperature set to 5 K there is a wide quench safety margin of about 25 K. For example, according to Eq. (1), $B$=1T field in $g$=40 mm gap can be achieved with the total transport current of $I·N$=36 kA that requires the superconductor cross-section of only about 1 cm$^2$ for the 30 K operations. That surely will be more than enough to operate at 5 K. Figure 2 shows the magnetic field simulation [26] for 1T 40 mm gap magnetic core made of Fe3%Si laminations with the HTS conductor contained within 80 mm (v) x 10 mm (h) space, i.e., with 8 cm$^2$ conductor cross-section area determined mostly by the size of the cooling liquid helium conduit pipes which support the HTS strands. To keep the AC losses in the HTS conductor at minimum, the conductor is placed such that the $B$-field crossing cable space is less than 5 % of the beam gap field $B$, i.e., less than 0.05 T.

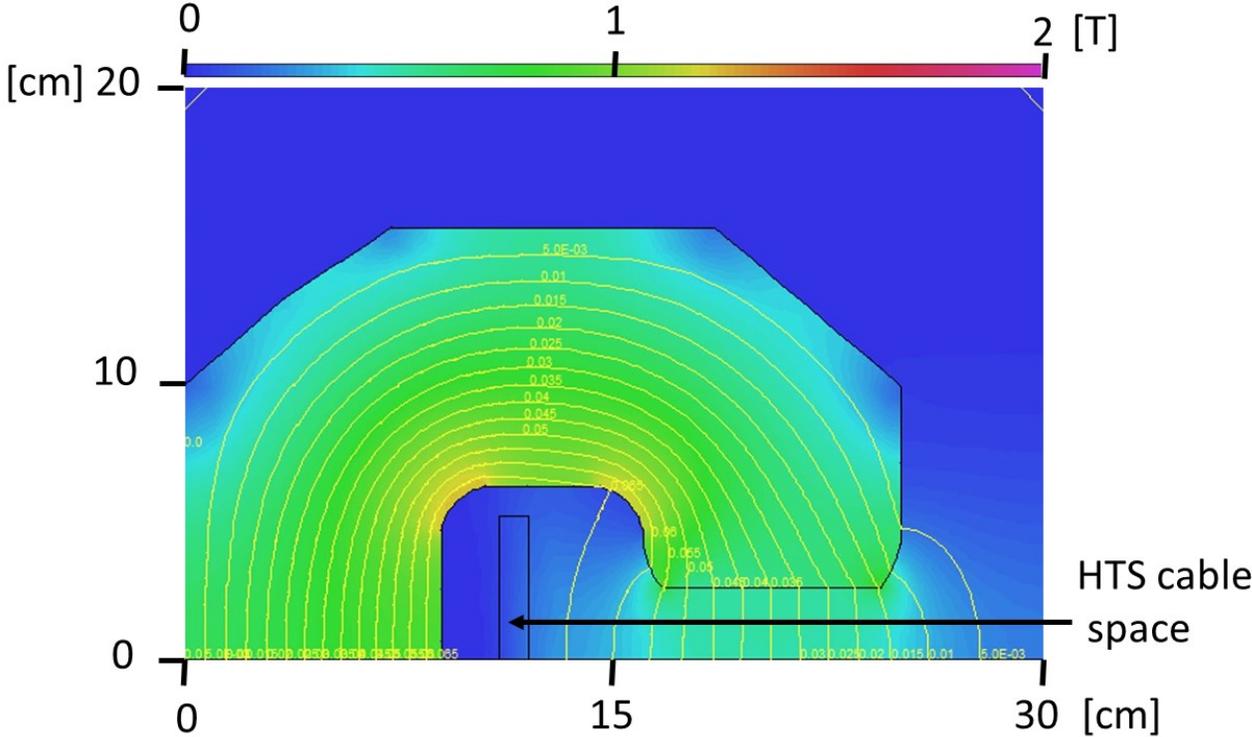

FIG. 2. Field simulations for 1 T magnet the 40 mm x 100 mm beam gap (quarter of the magnet shown).

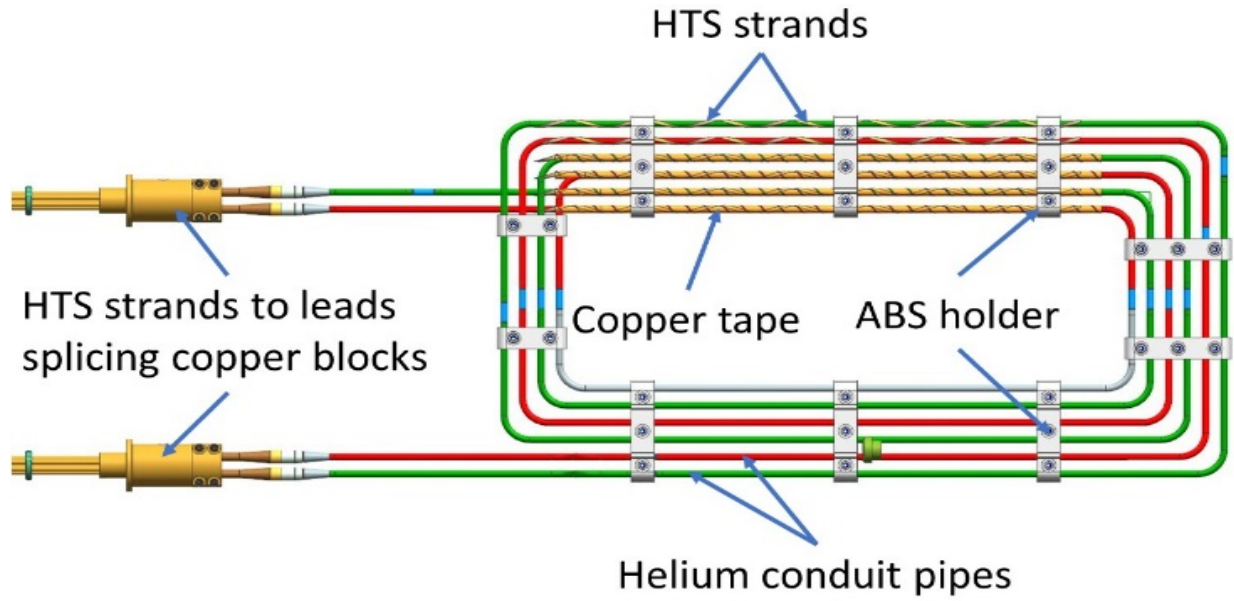

FIG. 3. Engineering design of the 3-turn HTS magnet power coil

The hysteresis power loss in the superconductor scales mostly linearly with the *dB/dt* rate [27]. The power losses in the magnetic core and in the resistive components of the cable structure are dominated by the eddy currents and scale as the $(dB/dt)^2$ and square of the thickness of these components [28]. With 100 μm laminations of the Fe3%Si prototype magnet core and the 0.5 mm thickness of SS cooling helium conduit pipes, these resistive power losses are strongly minimized and are negligible for the *dB/dt* rate range used in the test.

The 0.5 m long prototype magnet featuring two beam gaps of 10 mm (height) × 100 mm (width) is constructed of three parts allowing simple assembly and installation of the HTS power coil. The 3-turn conductor coil is shown in figure 3. Each coil turn consists of two helium conduit pipes. The 0.1 mm thick and 2 mm wide Re-BCO strands (SCS2050, Super-Power, Inc. [29]) are helically wound on the surface of the helium conduit pipes made of 316LN stainless steel, 8 mm OD, 0.5 mm wall thickness. Up to 12 Re-BCO strands can be placed on each helium conduit pipe. For the test magnet there are 2 strands attached to each conduit pipe for a total of 12 strands in the magnet power cable of the 24 m total length of which 12 m are inside the magnet core. The 0.1 mm thick, 12.5 mm wide oxygen-free high thermal conductivity copper tape is wound helically over the strands to firmly secure their attachment to the cooling helium conduit pipe.

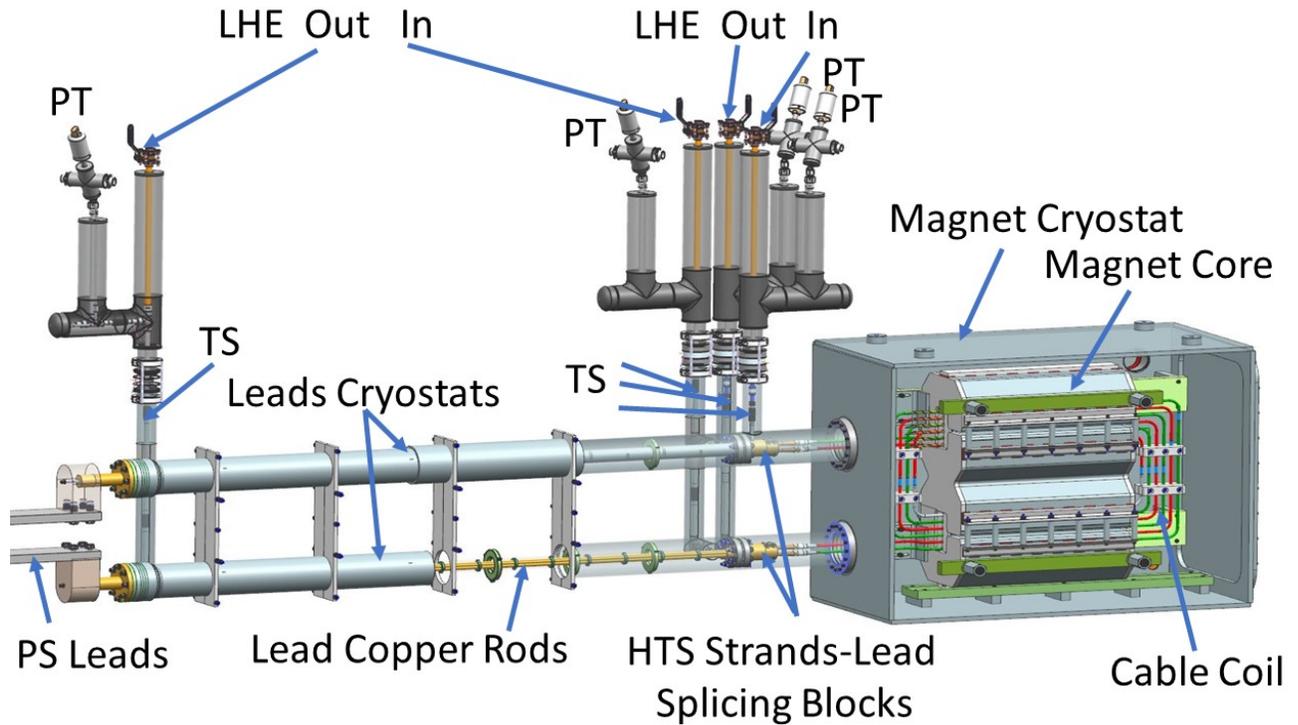

FIG. 4. The HTS magnet test system arrangement. TS - temperature sensors. PT - pressure transducers.

The test system arrangement is shown in figure 4. The magnet HTS conductor coil and the conventional (copper) current leads are cooled using separate liquid helium flows. The pressure and temperature sensors are placed at the inlets and outlets of the helium cooling circuits. The liquid helium from the magnet coil and the current leads exits into the 3.6 kW pipe-type heaters that warm liquid helium to the room temperature before passing it into the flow meters and then returning under suction to the cryogenic plant. The difference in helium temperatures measured at the inlet and outlet of the conductor helium conduit pipe together with measured helium pressure and flow rate were used to determine the cryogenic power loss through the change of helium enthalpy [30], $\Delta H = H$ (He out) $- H$ (He in). The change in enthalpy equals the change of system energy which when combined with helium flow rate ($F$) gives the generated heat: $Q$ [W] = $\Delta H$ (J/g) x $F$ (g/s). The 6.5 K single-phase liquid helium system of 1.7 bar pressure and 1 g/s flow rate was used for the conductor cooling. Although the Cernox temperature sensors [31] were calibrated with the precision of ±0.02 K the estimated inlet-outlet differential temperature readout uncertainty was ±0.1 K.

The projected prototype HTS magnet critical current is 6 kA at 30 K. But lack of stability of the supplied helium pressure has limited AC power supply operations to a maximum of $I$ = 2 kA current to avoid the HTS conductor quench during the low helium pressure (~ 1 bar) and high temperature (> 40 K) excursions. The AC current source is constructed of three, 1.5 Volt switcher cells similar to those reported in [32], arranged in series to maximize the output current. As the result of the test, operating the AC power supply at the 20 Hz repetition rate with $dI/dt$=38 kA/s current sine-wave, we obtained the $dB/dt$ rate of 12 T/s in each of the magnet's two gaps – see figure 5. At such $B$-field cycling rate no measurable helium temperature rise in the conductor loop was observed, therefore with the temperature sensors error of ±0.1 K we project the upper limit of about 0.8 W for the cryogenic power loss in the HTS magnet cable.

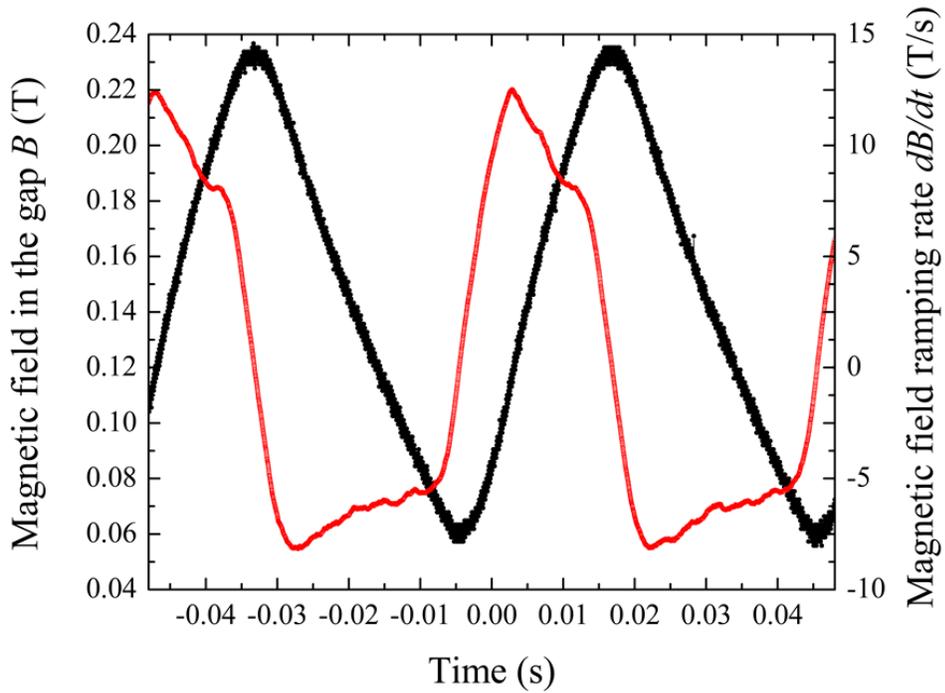

FIG. 5. Magnetic field $B$ (black line, left vertical axis) and its ramping rate $dB/dt$ (red line, right axis) in the HTS magnet gaps at 20 Hz.

In our previous study [33], the cryogenic power losses for the cable constructed of twenty 4.2 mm × 0.25 mm YBCO strands exposed to the ramping external fields of $dB/dt$= (4 - 20) T/s at 6.5 K were reliably measured. Also measured for comparison were the losses in the NbTi-based SC cable constructed to carry the same critical current at the same temperature. Figure 6 presents the results for both LTS and HTS cables and clearly indicates significant advantage of the latter. The YBCO-based cable data are for the strands with wide surface arranged approximately parallel to the magnetic field, aligned within the about 8º. It turned out that the NbTi cable data matched well the AC loss projections. The data for the YBCO-based cable show power losses significantly lower than those of the NbTi cable but exceeded projections by about a factor of two, possibly due to the twisting of the HTS strands inside the cryogenic pipe caused by the magnetic force. Under the conditions of the previous and current tests – relatively low *B*-fields and *dB/dt* rates - the hysteresis AC power loss in the superconductors are dominant mechanisms for the YBCO and the Re-BCO cables, so one can use the previous test results to estimate expected losses in the test described above. Using the ratio of the Re-BCO to the YBCO superconductor volumes (24 mm$^3$ / 192 mm$^3$) in the two tests and scaling to the ratio of the *B*-field ramping rates (1.2 T/s vs 12 T/s) one can project the cryogenic power loss to be about 0.06 W for the Re-BCO based cable, i.e., about an order of magnitude lower than the upper limit of 0.8 W obtained in the present test.  We plan to improve the AC power supply system to energize the Re-BCO based prototype magnet in the *dB/dt* range from 20 T/s up to 200 T/s which is required for, e.g., future muon accelerators [7,8] and expect to be able to determine the AC power loss more accurately combining the cryogenic and electric measurement methods [34,35].

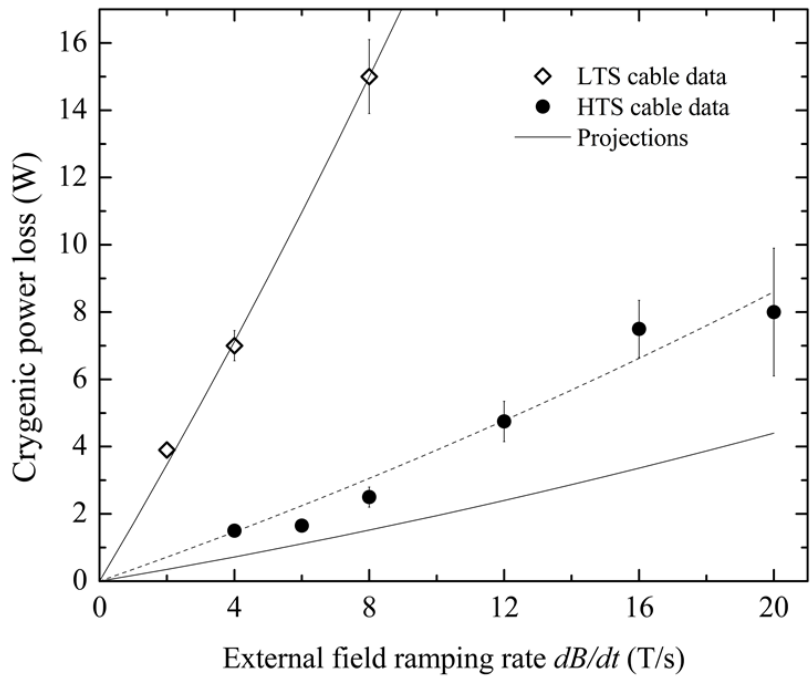

FIG. 6.  Cryogenic power losses measured for LTS and HTS cables (adapted from [33]).

In conclusion, we have developed the HTS-based fast cycling magnet suitable for a wide range of applications in high energy charged particle accelerators, have experimentally confirmed superiority of the HTS high-current conductors over traditional LTS ones in terms of much smaller AC losses in them, and finally, demonstrated the record high magnetic field ramping rates *dB/dt* of 12 T/s in the superconducting accelerator magnet prototype. Our results open new opportunities for the HTS magnet technology and for further developments toward required field quality, higher fields and other operationally critical properties of such magnets.

Authors would like to thank Frank McConologue for thoughtful engineering designs and Thomas Lynn for meticulous magnet assembly work. Fermilab is operated by Fermi Research Alliance, LLC under Contract No. DE-AC02-07CH11359 with the United States Department of Energy.